\documentclass[
    ,final            % use final for the camera ready runs
  ]
  {aipproc}
\layoutstyle{6x9}

\usepackage{graphicx}
\usepackage{amsmath}
\usepackage{color}

\definecolor{darkgreen}{rgb}{0,0.5,0}
\definecolor{purple}{rgb}{0.5,0,0.5}
\definecolor{nblue}{rgb}{0.0,0.0,0.50}
\definecolor{scarlet}{rgb}{1.0,0.2,0}

\newcommand{\lsim}{\mathrel{\rlap{\lower4pt\hbox{\hskip0pt$\sim$}}
\raise1pt\hbox{$<$}}}           %less than or approx. symbol
\newcommand{\gsim}{\mathrel{\rlap{\lower4pt\hbox{\hskip0pt$\sim$}}
\raise1pt\hbox{$>$}}}           %greater than or approx. symbol

\begin{document}

\title{Empirically Charting Dynamical Chiral Symmetry Breaking}


\author{Lei Chang}{
  address={Institute of Applied Physics and Computational Mathematics, Beijing 100094, China}
  }

\author{Craig D.\ Roberts}{
  address={Physics Division, Argonne National Laboratory, Argonne, Illinois 60439, USA},
  altaddress={Department of Physics, Peking University, Beijing 100871, China}
  }

%\begin{abstract}
%---Not to be made explicit:
%% We provide a snapshot of recent progress in hadron physics made using QCD's Dyson-Schwinger equations, reviewing the generation of a quark anomalous chromomagnetic moment, which may explain the longstanding puzzle of the $a_1$-$\rho$ mass splitting and the form of the pion and kaon valence-quark parton distribution functions.
%\end{abstract}

\maketitle

The goal of hadron physics is the provision of a quantitative explanation of the properties of hadrons through a solution of quantum chromodynamics (QCD).  This is a fundamental problem that is unique in the history of science: never before have we been confronted by a theory whose elementary excitations are not those degrees-of-freedom readily accessible via experiment.  Moreover, QCD generates forces which are so strong that less-than 2\% of a nucleon's mass can be attributed to the so-called current-quark masses that appear in the QCD Lagrangian; viz., forces that generate mass from almost nothing, a phenomenon known as dynamical chiral symmetry breaking (DCSB).

Neither confinement nor DCSB is apparent in QCD's Lagrangian and yet they play the dominant role in determining the observable characteristics of real-world QCD.  The physics of hadrons is ruled by \emph{emergent phenomena}, such as these, which can only be elucidated through the employment of nonperturbative methods in quantum field theory.  This is both the greatest novelty and the greatest challenge within the Standard Model.  We must find essentially new ways and means to explain precisely via mathematics the observable content of QCD.  Herein we provide a snapshot of recent progress in hadron physics made using QCD's Dyson-Schwinger equations (DSEs).  The complex of DSEs is a powerful tool, which has been employed with marked success to study confinement and DCSB, and their impact on hadron observables.  This is apparent from Ref.\,\cite{Chang:2010jq}.

Asymptotic coloured states have not been observed, so no solution to QCD will be complete if it does not explain confinement.  This means confinement in the real world, which contains quarks with light current-quark masses.  That is distinct from the artificial universe of pure-gauge QCD without dynamical quarks, studies of which tend merely to focus on achieving an area law for a Wilson loop and hence are irrelevant to the question of light-quark confinement.

Confinement can be related to the analytic properties of QCD's Schwinger functions \cite{Krein:1990sf} and can therefore be translated into the challenge of charting the infrared behavior of QCD's \emph{universal} $\beta$-function.  This is a well-posed problem whose solution can be addressed in any framework enabling the nonperturbative evaluation of renormalisation constants.
In this connection we note that the hadron spectrum \cite{Holl:2005vu}, and hadron elastic and transition form factors \cite{Cloet:2008re,Aznauryan:2009da} provide unique information about the long-range interaction between light-quarks and, in addition, the distribution of a hadron's characterising properties amongst its QCD constituents.  However, to make full use of extant and forthcoming data, it will be necessary to have Poincar\'e covariant theoretical tools that enable the reliable study of hadrons in the mass range $1$-$2\,$GeV.  Crucially, on this domain both confinement and DCSB are germane; and the DSEs provide such a tool.

DCSB; namely, the generation of mass \emph{from nothing}, does take place in QCD.  It arises primarily because a dense cloud of gluons comes to clothe a low-momentum quark.  This is best seen by solving the DSE for the dressed-quark propagator \cite{Bhagwat:2007vx}; i.e., the gap equation.  However, the origin of the interaction strength at infrared momenta, which guarantees DCSB through the gap equation, is unknown.  This relationship ties confinement to DCSB.  The reality of DCSB means that the Higgs mechanism is largely irrelevant to the bulk of normal matter in the universe.  Instead the single most important mass generating mechanism for light-quark hadrons is the strong interaction effect of DCSB; e.g., one can identify it as being responsible for 98\% of a proton's mass.

In chiral-limit QCD, DCSB is most basically expressed in a strongly momentum-dependent dressed-quark mass; viz., $M(p^2)$ in the quark propagator:
\begin{equation}
S(p) = \frac{1}{i \gamma\cdot p A(p^2) + B(p^2)} = \frac{Z(p^2)}{i \gamma\cdot p + M(p^2)}\,.
\end{equation}
The appearance and behaviour of $M(p^2)$ are essentially quantum field theoretic effects, unrealisable in quantum mechanics.  The running mass connects the infrared and ultraviolet regimes of the theory, and establishes that the constituent-quark and current-quark masses are simply two connected points on a single curve separated by a large momentum interval.  QCD's dressed-quark behaves as a constituent-quark, a current-quark, or something in between, depending on the momentum of the probe which explores the bound-state containing the dressed-quark.  It follows that calculations addressing momentum transfers $Q^2 \gsim M^2$, where $M$ is the mass of the hadron involved, require a Poincar\'e-covariant approach that can veraciously realise quantum field theoretical effects \cite{Cloet:2008re}.  Owing to the vector-exchange character of QCD, covariance also guarantees the existence of nonzero quark orbital angular momentum in a hadron's rest-frame \cite{Roberts:2007ji,Bhagwat:2006xi}.

Through the gap and Bethe-Salpeter equations (BSEs) the pointwise behaviour of QCD's $\beta$-function determines the pattern of chiral symmetry breaking.  Since these and other DSEs connect the $\beta$-function to experimental observables, the comparison between computations and observations of the hadron properties can be used to chart the
$\beta$-function's long-range behaviour.  In order to realise this goal a nonperturbative symmetry-preserving DSE truncation is necessary.  Steady progress has been made with a scheme that is systematically improvable \cite{Munczek:1994zz,Bender:1996bb}.  On the other hand, significant qualitative advances in understanding the essence of QCD could be made with symmetry-preserving kernel \emph{Ans\"atze} that express important additional nonperturbative effects, which are impossible to capture in any finite sum of contributions.

Such an approach is now available \cite{Chang:2009zb}.  It begins with a novel form for the axial-vector BSE, which is valid when the quark-gluon vertex is fully dressed.  Therefrom, a Ward-Takahashi identity for the Bethe-Salpeter kernel is derived and solved for a class of dressed quark-gluon-vertex models.  The solution yields a symmetry-preserving closed system of gap and vertex equations.  As the analysis can readily be extended to the vector equation, a comparison is possible between the responses of pseudoscalar- and scalar meson masses to nonperturbatively dressing the quark-gluon vertex.  The result indicates that spin-orbit splitting in the meson spectrum is enormously enhanced by DCSB.

It has been conjectured \cite{Chang:2009zb} that the full realisation of DCSB in the Bethe-Salpeter kernel will have a material impact on mesons with mass greater than 1\,GeV.  Moreover, that it can overcome a longstanding failure of theoretical hadron physics.  Namely, no extant hadron spectrum calculation is believable because all symmetry preserving studies produce a splitting between vector and axial-vector mesons that is far too small: just one-quarter of the experimental value (see, e.g., Refs.\,\cite{Watson:2004kd,Maris:2006ea,Fischer:2009jm}).  In this connection, preliminary results %\footnote{Preliminary in the sense that not all Dirac covariants are employed in solving for the vertices and internal consistency checks for the pseudoscalar and scalar channels are still underway.}
are now available \cite{Chang:2010jq} and they are listed in Table~\ref{massresult}.  The second numerical column reports results obtained in rainbow-ladder truncation; viz., leading-order in the systematic and symmetry-preserving truncation scheme of Ref.\,\cite{Bender:1996bb}.  As anticipated, while the $\rho$-meson mass is acceptable, the $a_1$-mass is far too small.

\begin{table}[t]
\begin{tabular}{lcccc}
\hline
  & \tablehead{1}{c}{b}{experiment\\}
  & \tablehead{1}{c}{b}{rainbow-ladder\\}
  & \tablehead{1}{c}{b}{Ball-Chiu consistent\\}
  & \tablehead{1}{c}{b}{Ball-Chiu \\ plus anom.\ cm mom.}\\
  \hline
mass $a_1$ & 1230 & 759 & 1066 & 1230    \\
mass $\rho$& ~775& 644 & ~924 & ~745  \\
mass splitting & ~455 & 115 & ~142 & ~485  \\
\hline
\end{tabular}
\caption{\label{massresult}
Axial-vector and vector meson masses calculated in three truncations of the coupled gap and Bethe-Salpeter equations.  The last column was obtained using the standard Ball-Chiu \emph{Ansatz} augmented by the quark anomalous chromomagnetic moment in Eqs.\,(\protect\ref{qcdanom1}), (\protect\ref{qcdanom2}).}
\end{table}

The procedure introduced in Ref.\,\cite{Chang:2009zb} enables meson masses to be calculated using a symmetry-preserving DSE truncation whose diagrammatic content is unknown.  One can therefore elucidate the effect of an essentially nonperturbative DCSB component in dressed-quark gluon vertex on the $\rho$-$a_1$ complex; in this case, the $\Delta_B$ term in the Ball-Chiu vertex \cite{Ball:1980ay}, which had an enormous impact in the scalar channel:
\begin{equation}
i\Gamma_\mu(q,k)  =
i\Sigma_{A}(q^2,k^2)\,\gamma_\mu +
2 \ell_\mu \left[i\gamma\cdot \ell \,
\Delta_{A}(q^2,k^2) + \Delta_{B}(q^2,k^2)\right] \!,
\label{bcvtx}
\end{equation}
where $\ell=(q+k)/2$, $\Sigma_{\Phi}(q,k;P) = [\Phi(q;P)+\Phi(k;P)]/2$ and $\Delta_{\Phi}(q,k;P) = [\Phi(q;P)-\Phi(k;P)]/[q^2-k^2]$.  The results obtained with this vertex are shown in Table~\ref{massresult}: the DCSB $\Delta_B$-term boosts the $a_1$ mass, which is a positive outcome, but it simultaneously boosts the $\rho$ mass, such that the mass-splitting is practically unchanged from the rainbow-ladder result.  Was Ref.\,\cite{Chang:2009zb} too optimistic in imagining that the new scheme could provide the first realistic meson spectrum encompassing states with mass greater-than 1\,GeV?

Before answering, let us return to a consideration of chirally symmetric QCD.  That theory exhibits helicity conservation so that, perturbatively, the quark-gluon vertex cannot have a term with the helicity-flipping characteristics of $\Delta_B$.  There is another feature of massless fermions in gauge field theories; namely, they cannot posses an anomalous chromo/electro-magnetic moment because the term that describes it couples left- and right-handed fermions.  However, if the theory's chiral symmetry is strongly broken dynamically, why shouldn't the fermions posses a large anomalous chromo/electro-magnetic moment?  Such an effect is expressed in the quark-gluon-vertex via a term
\begin{equation}
\label{qcdanom1}
\Gamma_\mu^{\rm acm} (q,k) = \sigma_{\mu\nu} (q-k)_\nu \, \tau_5(q,k)
\end{equation}
where, owing to DCSB, the natural strength is represented by the \emph{Ansatz}
\begin{equation}
\label{qcdanom2}
\tau_5(q,k) = \Delta_B(q^2,k^2)\,.
\end{equation}
NB.\ Based on the models in Refs.\,\cite{Cloet:2008re,Ivanov:2007cw}, $2 M_E Z(M_E^2) \Delta_B(M_E^2,M_E^2) \sim - \frac{1}{2}$, where $M_E$ is the Euclidean constituent-quark mass, defined in Ref.\,\cite{Maris:1997tm}.

Using the procedure introduced in Ref.\,\cite{Chang:2009zb}, the vector and axial-vector vertex equations can be solved using the dressed-quark-gluon vertex obtained as the sum of Eqs.\,(\ref{bcvtx}) and (\ref{qcdanom1}).  The effect is remarkable: the anomalous chromomagnetic moment leads to additional repulsion in the $a_1$ channel but significant attraction in the $\rho$ channel such that, for the first time, a realistic result is simultaneously obtained for the masses in both these channels, and hence the $a_1$-$\rho$ mass-splitting -- see Table~\ref{massresult}.
Furthermore, the origin of the splitting, in an interference between $\Delta_B$ in Eq.\,(\ref{bcvtx}) and $\tau_5$ in Eqs.\,(\protect\ref{qcdanom1}), (\protect\ref{qcdanom2}) is intuitively appealing.  In the chiral limit the mass-splitting between parity partners should owe solely to DCSB and here that is seen explicitly: in the absence of DCSB, $\Delta_B\equiv 0 \equiv \tau_5$.  The rainbow-ladder result is also understood: this truncation fails to adequately express DCSB in the Bethe-Salpeter kernel and hence cannot realistically split parity partners.

Table~\ref{massresult} is a ``first-guess'' result; i.e., there was no tuning of the strength in Eq.\,(\ref{qcdanom2}), so how reliable can it be?  This question amounts to deciding whether a realistic size is assumed for a light-quark's anomalous chromomagnetic moment.  Fortunately, an analysis is available of results for the dressed-quark-gluon vertex obtained through numerical simulations of quenched-QCD \cite{Skullerud:2003qu}.  This study shows that $\tau_5$ is dynamically two orders-of-magnitude larger than the one-loop perturbative result and, indeed, is of the same magnitude and possesses the same domain of significant support as $\Delta_B(q^2,k^2)$, precisely in accordance with the assumption we have made.  In addition, efficacious model studies support a moment of similar size \cite{Kochelev:1996pv,Ebert:2005es}.

At this point it is natural to consider whether DCSB in QCD can also generate a large quark anomalous \emph{electro}magnetic moment term in the quark-photon vertex.  In perturbation theory, of course, since it doesn't express DCSB, the quark's anomalous electromagnetic moment is small \cite{Bekavac:2009wh}.  One obtains the same answer in the rainbow-ladder truncation; e.g., the $F_6$ and $F_8$ terms in Ref.\,\cite{Maris:1999bh}, which combine to form $\tau_5$ in our notation, contribute less-than 1\% to the pion's electromagnetic form factor.  However, as we've already seen, this truncation doesn't adequately incorporate DCSB into the Bethe-Salpeter kernel.  Only the method of Ref.\,\cite{Chang:2009zb} can readily provide an answer to this question and we are currently compiling the necessary information.

Another important problem is the computation of the parton distribution functions of the most accessible hadrons \cite{Holt:2010vj}.  In connection with uncovering the essence of the strong interaction, the behaviour of the valence-quark distribution functions at large Bjorken-$x$ is most relevant.  Owing to the dichotomous nature of Goldstone bosons, understanding the valence-quark distribution functions in the pion and kaon is of great importance.  Moreover, given the large value of the ratio of $s$-to-$u$ current-quark masses, a comparison between the pion and kaon structure functions offers the chance to chart effects of explicit chiral symmetry breaking on the structure of would-be Goldstone modes.  There is also the prediction \cite{Ezawa:1974wm,Farrar:1975yb} that a theory in which the quarks interact via $1/k^2$ vector-boson exchange will produce valence-quark distribution functions for which
\begin{equation}
q_{\rm v}(x) \propto (1-x)^{2+\gamma} \,,\; x\gsim 0.85\,,
\end{equation}
where $\gamma\gsim 0$ is an anomalous dimension that grows with increasing momentum transfer.  (See Sec.VI.B.3 of Ref.\,\cite{Holt:2010vj} for a detailed discussion.)

Experimental knowledge of the parton structure of the pion and kaon arises primarily from pionic or kaonic Drell-Yan scattering from nucleons in heavy nuclei \cite{Badier:1980jq,Wijesooriya:2005ir}.  Theoretically, given that DCSB plays a crucial role in connection with pseudoscalar mesons, one must employ an approach that realistically expresses this phenomenon.  The DSEs therefore provide a natural framework: a study of the pion exists \cite{Hecht:2000xa} and one of the kaon is underway \cite{kaonpdf}.

\begin{figure}[t]
\vspace*{-7ex}

\includegraphics[clip,width=0.33\textheight]{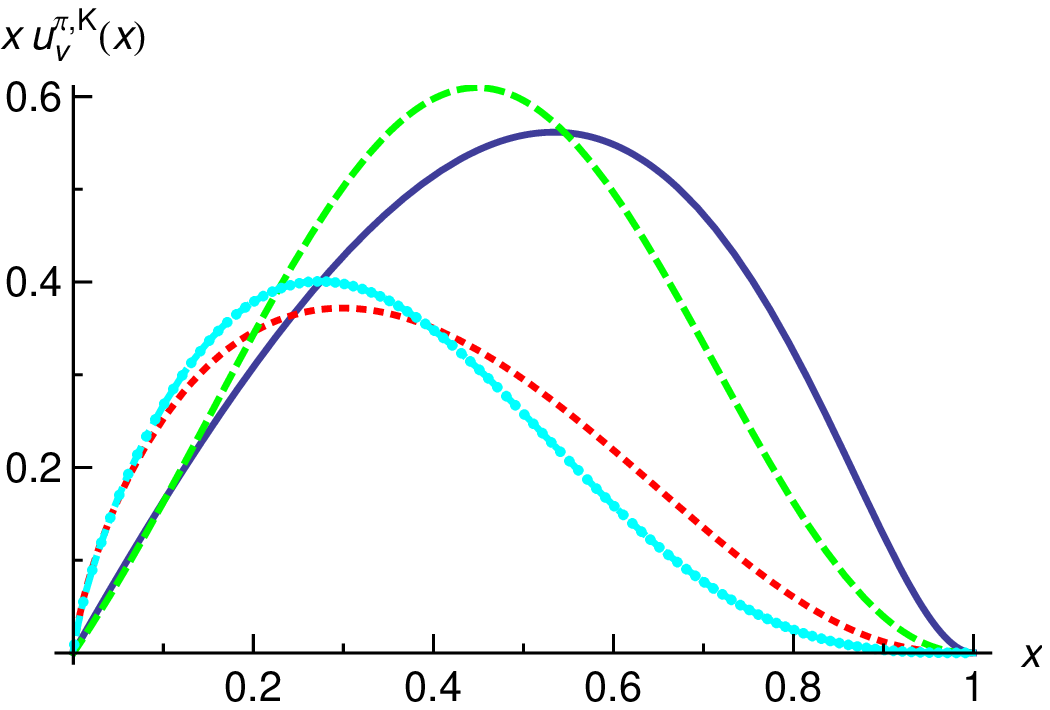}\hspace*{0.5em}
\vspace*{-7ex}

\includegraphics[clip,width=0.33\textheight]{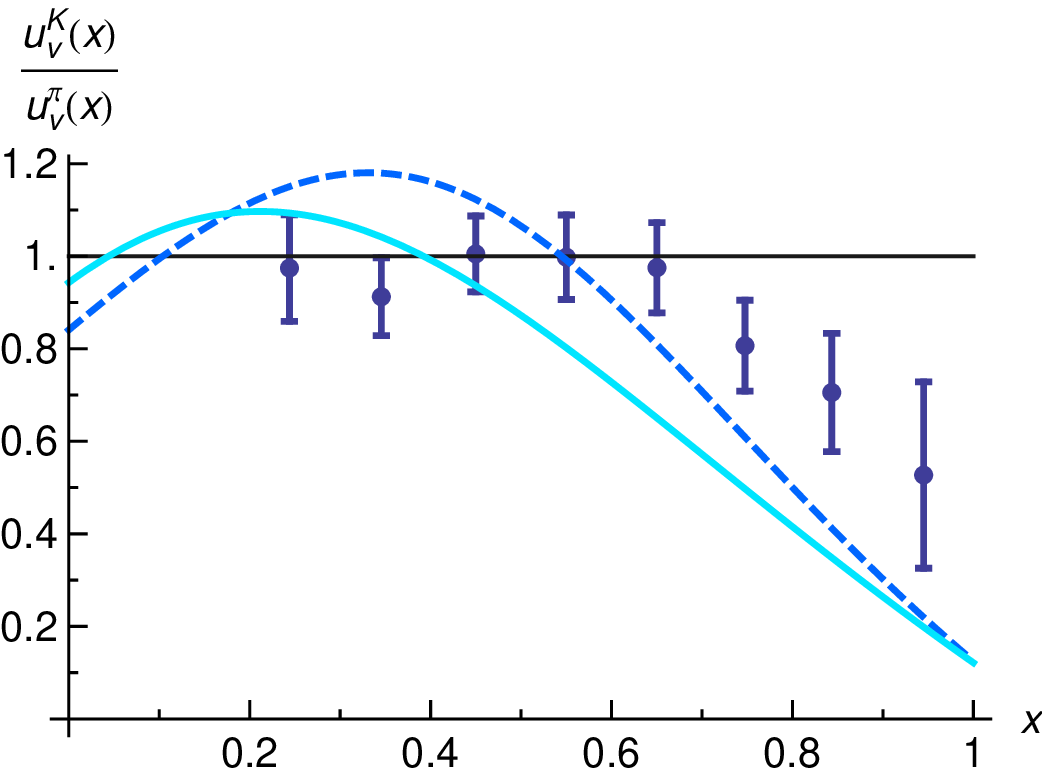}
\vspace*{-5ex}

\caption{\label{ShigetaniSoft}
\underline{Left panel}. Valence $u$-quark distribution functions, computed as described in the text: \emph{solid curve}, $u_{\rm v}^\pi(x)$; and \emph{dashed curve}, $u_{\rm v}^K(x)$.  Applying leading-order QCD evolution from $Q_0^2=0.32\,$GeV$^2$ to $Q^2=25\,$GeV$^2$, explained in Sec.\,II.D of Ref.\,\protect\cite{Holt:2010vj}, one obtains the other two curves from these starting distributions: \emph{dashed curve}, $u_{\rm v}^\pi(x;25)$; and \emph{dotted curve}, $u_{\rm v}^K(x,25)$.
\underline{Right panel}.  Model ratio $u_{\rm v}^K/u_{\rm v}^\pi$ evaluated at $Q_0^2=0.32\,$GeV$^2$, \emph{Dashed curve}, and $Q^2=25\,$GeV$^2$, \emph{solid curve}. $\left. u_{\rm v}^K/u_{\rm v}^\pi\right|_{x=1}=0.13$.  Under the right conditions, $u_{\rm v}^K/u_{\rm v}^\pi$ should equal the ratio of kaon-to-pion Drell-Yan cross-sections, and we reproduce that obtained from a sample of dimuon events with invariant mass $4.1<M<8.5\,$GeV \protect\cite{Badier:1980jq}.}
\end{figure}

The results to be anticipated from the latter study have been illustrated \cite{Chang:2010jq} through an internally consistent calculation based upon the QCD-improvement of a simple model used already for pion and kaon distribution functions \cite{Shigetani:1993dx}.
In Fig.\,\ref{ShigetaniSoft} we depict our computed distributions themselves and relevant ratios.  Aspects of the curves are model-independent.
For example, owing to its larger mass, one anticipates that the $s$-quark should carry more of the charged-kaon's momentum than the $u$-quark.  This explains why the support of $x u_{\rm v}^K(x)$ is shifted to lower-$x$ than that $x u_{\rm v}^\pi(x)$.
QCD evolution is an area-conserving operation on the distribution function, which shifts support from large-$x$ to small-$x$.  Thus, while both $u_{\rm v}^{\pi,K}(x;Q_0) \propto (1-x)^2$ for $x\simeq 1$,
\begin{equation}
u_{\rm v}^{\pi,K}(x;Q) \stackrel{x\simeq 1}{\propto} (1-x)^a, \; a= 2.7\,.
\end{equation}
These observations explain the qualitative behaviour of the evolved distributions.
Concerning the ratio, as a consequence of the form of the evolution equations its value at $x=1$ is invariant under evolution, whereas the value at $x=0$ approaches one under evolution owing to the increasingly large population of sea-quarks produced thereby.

There are many reasons why this is an exciting time in hadron physics.  We have focused on one.  Namely, through the DSEs, we are positioned to unify phenomena as apparently diverse as the: hadron spectrum; hadron elastic and transition form factors, from small- to large-$Q^2$; and parton distribution functions.  The key is an understanding of both the fundamental origin of nuclear mass and the far-reaching consequences of the mechanism responsible; namely, DCSB.  These things might lead us to an explanation of confinement, the phenomenon that makes nonperturbative QCD the most interesting piece of the Standard Model.

%%%%%%%%%%%%%%%%%%%%%%%%%%%%%%%%%%%%%%%%%%%%
%% MAINMATTER
%%%%%%%%%%%%%%%%%%%%%%%%%%%%%%%%%%%%%%%%%%%%

%%%%%%%%%%%%%%%%%%%%%%%%%%%%%%%%%%%%%%%%%%%%%%%%
%% BACKMATTER
%%%%%%%%%%%%%%%%%%%%%%%%%%%%%%%%%%%%%%%%%%%%%%%%

\bigskip

%\begin{theacknowledgments}
\hspace*{-\parindent}\mbox{\textbf{Acknowledgments.}}~Work supported by:
the National Natural Science Foundation of China, contract no.\ 10705002;
the U.\,S.\ Dept.\ of Energy, Office of Nuclear Physics, contract no.~DE-AC02-06CH11357;
and the U.\,S.\ National Science Foundation, grant no.\ PHY-0903991, in conjunction with a CONACyT Mexico-USA collaboration grant.
%\end{theacknowledgments}

\vspace*{-2ex}

%%%%%%%%%%%%%%%%%%%%%%%%%%%%%%%%%%%%%%%%%%%%%%%%
%% The bibliography can be prepared using the BibTeX program or
%% manually.
%%
%% The code below assumes that BibTeX is used.  If the bibliography is
%% produced without BibTeX comment out the following lines and see the
%% aipguide.pdf for further information.
%%
%% For your convenience a manually coded example is appended
%% after the \end{document}
%%%%%%%%%%%%%%%%%%%%%%%%%%%%%%%%%%%%%%%%%%%%%%%%

%%%%%%%%%%%%%%%%%%%%%%%%%%%%%%%%%%%%%%%%%%%%%%%%
%% You may have to change the BibTeX style below, depending on your
%% setup or preferences.
%%
%%
%% For The AIP proceedings layouts use either
%%%%%%%%%%%%%%%%%%%%%%%%%%%%%%%%%%%%%%%%%%%%

\bibliographystyle{aipproc}   % if natbib is available
%\bibliographystyle{aipprocl} % if natbib is missing

%%%%%%%%%%%%%%%%%%%%%%%%%%%%%%%%%%%%%%%%%%%
%% You probably want to use your own bibtex database here
%%%%%%%%%%%%%%%%%%%%%%%%%%%%%%%%%%%%%%%%%%%
%\bibliography{sample}

%%%%%%%%%%%%%%%%%%%%%%%%%%%%%%%%%%%%%%%%%%%
%% Just a reminder that you may have to run bibtex
%% All of it up to \end{document} can be removed
%% if you don't like the warning.
%%%%%%%%%%%%%%%%%%%%%%%%%%%%%%%%%%%%%%%%%%%
\IfFileExists{\jobname.bbl}{}
 {\typeout{}
  \typeout{******************************************}
  \typeout{** Please run "bibtex \jobname" to optain}
  \typeout{** the bibliography and then re-run LaTeX}
  \typeout{** twice to fix the references!}
  \typeout{******************************************}
  \typeout{}
 }

%%%%%%%%%%%%%%%%%%%%%%%%%%%%%%%%%%%%%%%%%%%
%% The following lines show an example how to produce a bibliography
%% without the help of the BibTeX program. This could be used instead
%% of the above.
%%%%%%%%%%%%%%%%%%%%%%%%%%%%%%%%%%%%%%%%%%%

\end{document}